# Similarity Join Size Estimation using Locality Sensitive Hashing[*]

Hongrae Lee
University of British Columbia
xguy@cs.ubc.ca

Raymond T. Ng
University of British Columbia
rng@cs.ubc.ca

Kyuseok Shim
Seoul National University
shim@ee.snu.ac.kr

## ABSTRACT

Similarity joins are important operations with a broad range of applications. In this paper, we study the problem of vector similarity join size estimation (VSJ). It is a generalization of the previously studied set similarity join size estimation (SSJ) problem and can handle more interesting cases such as TF-IDF vectors. One of the key challenges in similarity join size estimation is that the join size can change dramatically depending on the input similarity threshold.

We propose a sampling based algorithm that uses Locality-Sensitive-Hashing (LSH). The proposed algorithm *LSH-SS* uses an LSH index to enable effective sampling even at high thresholds. We compare the proposed technique with random sampling and the state-of-the-art technique for SSJ (adapted to VSJ) and demonstrate *LSH-SS* offers more accurate estimates throughout the similarity threshold range and small variance using real-world data sets.

## 1. INTRODUCTION

Given a similarity measure and a minimum similarity threshold, a *similarity join* is to find all pairs of objects whose similarity is not smaller than the similarity threshold. The object in a similarity join is often a vector. For instance, a document can be represented by a vector of words in the document, or an image can be represented by a vector from its color histogram. In this paper, we focus on the *vector* representation of objects and study the following problem.

DEFINITION 1 (THE VSJ PROBLEM). *Given a collection of real-valued vectors $V = \{v_1, ..., v_n\}$ and a threshold $\tau$ on a similarity measure sim, estimate the number of pairs $J = |\{(u,v) : u,v \in V, sim(u,v) \geq \tau, u \neq v\}|$.*

Similarity joins have a broad range of applications including near duplicate document detection and elimination,

[*]This work was supported by the National Research Foundation of Korea (NRF) grant funded by the Korea government (MEST) (No. 2010-0000793).

query refinement for web search, coalition detection [3], and data cleaning processes [17, 2]. Accordingly, similarity joins have recently received much attention, e.g. [17, 2, 6, 3, 11]. Chaudhuri et al. identified a similarity join operation as a primitive operator in database systems [6].

To successfully incorporate similarity join operations in database systems, it is imperative that we have reliable size estimation technique for them. The query optimizer needs accurate size estimations to produce an optimized query plan. Thus, in this paper, we focus on the size estimation of vector similarity joins.

In the literature, the similarity join size estimation problem has been defined using *sets* as follows:

DEFINITION 2 (THE SSJ PROBLEM). *Given a collection of real-valued sets $S = \{s_1, ..., s_n\}$ and a threshold $\tau$ on a similarity measure sim, estimate the number of pairs $J = |\{(r,s) : r,s \in S, sim(r,s) \geq \tau, r \neq s\}|$.*

Note that our formulation of similarity joins with vectors is more general and can handle more practical applications. For instance, while in the SSJ problem a document is simply a set of words in the document, in the VSJ problem a document can be modeled with a vector of words with TF-IDF weights. It can also deal with multiset semantics with occurrences. In fact, most of the studies on similarity joins first formulate the problem with sets and then extend it with TF-IDF weights, which is indeed a vector similarity join.

The SSJ problem has been previously studied by Lee et al. [14]. A straightforward extension of SSJ techniques for the VSJ problem is to embed a vector into a set space. We convert a vector into a set by treating a dimension as an element and repeating the element as many times as the dimension value, using standard rounding techniques if values are not integral [2]. In practice, however, this embedding can have adverse effects on performance, accuracy or required resources. Intuitively, a set is a special case of a binary vector and is not more difficult to handle than a vector. For instance, Bayardo et al. [3] define the vector similarity join problem and add special optimizations that are possible when vectors are binary vectors (sets).

In our VSJ problem, we consider cosine similarity as the similarity measure *sim* since it has been successfully used across several domains [3]. Let $u[i]$ denote the $i$-th dimension value of vector $u$. Cosine similarity is defined as $cos(u,v) = u \cdot v / \|u\| \|v\|$, where $u \cdot v = \sum_i u[i] \cdot v[i]$ and $\|u\| = \sqrt{\sum_i u[i]^2}$. We focus on self-joins and discuss extensions to general joins in Appendix B.2.

One of the key challenges in similarity join size estimation is that the join size can change dramatically depending on the input similarity threshold. While the join size can be close to $n^2$ at low thresholds where $n$ is the database size, it can be very small at high thresholds. For instance, in the DBLP data set, the join selectivity is only about 0.00001 % at $\tau = 0.9$. While many sampling algorithms have been proposed for the (equi-)join size estimation, their guarantees fail in such a high selectivity range, e.g. [15, 10, 9]. Intuitively, it is not practical to apply simple random sampling when the selectivity is very high. This is problematic since similarity thresholds between 0.5 and 0.9 are typically used [3]. Note that the join size in that range may be large enough to affect query optimization due to the large cross product size. Moreover, as observed in [13], join size errors propagate. That is, even if the original errors are small, their transitive effect can be devastating.

In this paper, we propose sampling based techniques that exploit the Locality Sensitive Hashing (LSH) scheme, which has been successfully applied in similarity searches across many domains. LSH builds hash tables such that similar objects are more likely to be in the same bucket. Our key idea is that although sampling a pair satisfying a high threshold is very difficult, it is relatively easy to sample the pair using the LSH scheme because it groups similar objects together. We show that the proposed algorithm $LSH\text{-}SS$ gives good estimates throughout the similarith threshold range with a sample size of $\Omega(n)$ pairs of vectors (i.e. $\Omega(\sqrt{n})$ tuples from each join relation in an equi-join) with probabilistic guarantees. The proposed solution only needs minimal addition to the existing LSH index and thus is readily applicable to many similarity search applications. As a summary, we make the following contributions:

- We present two baseline methods in Section 3. We consider random sampling and adapt Lattice Counting (LC) [14] which is proposed for the SSJ problem.

- We extend the LSH index to support similarity join size estimation in Section 4. We also propose $LSH\text{-}S$ which relies on an LSH function analysis.

- We describe a stratified sampling algorithm $LSH\text{-}SS$ that exploits the LSH index in Section 5. We apply different sampling procedures for the two partitions induced by an LSH index: pairs of vectors that are in the same bucket and those that are not.

- We compare the proposed solutions with random sampling and LC using real-world data sets in Section 6. The experimental results show that $LSH\text{-}SS$ is the most accurate with small variance.

## 2. RELATED WORK

Many algorithms have been proposed on the similarity join processing, e.g. [17, 2, 6, 3]. Their focus is not on size estimation. In fact, the existence of many join processing algorithms motivates the size estimation study for query optimization. Most of them studied set similarity joins with complex weights such as TF-IDF or vector similarity joins.

Lee et al. proposed LC for the SSJ problem [14]. We present this technique in more details in the following section and compare with the proposed solution in the experiments. Hadjieleftheriou et al. studied the problem of selectivity estimation for set similarity selection (not join) queries [11].

There have been many studies using random sampling for the (equi-)join size estimation; some examples include adaptive sampling [15], cross/index/tuple sampling [10], bifocal sampling [9], and tug-of-war [1]. Some of them can be adapted to similarity joins, but their guarantees do not hold due to differences in sampling cost models or they require impractical sample sizes. Below, we outline two closely related sampling algorithms.

Adaptive sampling is proposed by Lipton et al [15]. Its main idea is to terminate sampling process when the query size (the accumulated answer size from the sample so far) reaches a threshold not when the number of samples reaches a threshold. While it does not produce reliable estimates in skewed data, we observe that its adaptive nature can still be useful for the VSJ problem. It is used as a subroutine in our solution in Section 5.

Bifocal sampling is proposed by Ganguly et al. to cope with the skewed data problem [9]. It tackles the problem by considering high-frequency values and low-frequency values with separate procedures. However, as will be shown shortly, it cannot guarantee good estimates at high thresholds when applied to the VSJ problem.

## 3. BASELINE METHODS

### 3.1 Random Sampling

The first baseline method is uniform random sampling. We select $m$ pairs of vectors uniformly at random (with replacement) and count the number of pairs satisfying the similarity threshold $\tau$. We return the count scaled up by $M/m$ where $M$ denotes the total number of pairs of vectors in the database $V$. We also consider an alternative method of first sampling $\sqrt{m}$ records and computing similarities of all the pairs in the sample (cross sampling [10]).

Equi-join size estimation techniques mostly do not offer clear benefits over the simple random sampling in the VSJ problem. We note two challenges in the VSJ problem compared to the equi-join size estimation. In the equi-join size of $|R \bowtie S|$, we can focus on frequency distribution on the join column of each relation $R$ and $S$. For instance, if we know a value $v$ appears $n_r(v)$ times in $R$ and $n_s(v)$ times in $S$, the contribution of $v$ in the join size is simply $n_r(v) \cdot n_s(v)$, i.e. multiplication of two frequencies. We do not need to compare all the $n_r(v) \cdot n_s(v)$ pairs. In similarity joins, however, we need to actually compare the pairs to measure similarity. This difficulty invalidates the use of popular auxiliary structures such as indexes [10, 7] or histograms [7].

Furthermore, similarity join size at high thresholds can be much smaller than the join size assumed in equi-joins. For instance, in the DBLP data set ($n = 800K$), the join size of $\Omega(n \log n)$ assumed in bifocal sampling is more than $15M$ pairs and corresponds to the cosine similarity value of only about 0.4. In the most cases, users will be interested in much smaller join sizes and thus higher thresholds.

### 3.2 Adaptation of Lattice Counting

Lattice Counting (LC) is proposed by Lee et al. [14] to estimate SSJ size with Jaccard similarity. Jaccard similarity between two sets $A, B$ is defined as $sim_J(A, B) = |A \cap B|/|A \cup B|$. LC relies on succinct representation of sets using Min-Hashing [4]. A useful property of Min-Hashing is that if $h$ is a Min-Hash function then $P(h(A) = h(B)) = sim_J(A, B)$. For each set $A$, a signature of the set $sig(A)$ is constructed by concatenating $k$ Min-Hash functions. Jaccard similarity between two sets $A, B$ can be estimated by the number of positions of $sig(A)$ and $sig(B)$ which overlap. LC performs an analysis on the signatures of all sets. For our purposes, LC can be treated as a black box.

We observe that the analysis of LC is valid as long as the number of matching positions in the signatures of two objects is proportional to their similarity. Note that this requirement is exactly the property of the LSH scheme. Thus LC can be applied for the VSJ problem with an appropriate LSH scheme. In fact, Min-Hashing is the LSH scheme for Jaccard similarity. For the VSJ problem, we first build the signature database by applying an LSH scheme to the vector database and then apply LC.

## 4. LSH INDEX FOR THE VSJ PROBLEM

We first describe how we extend the LSH index and present a naive method with a uniformity assumption, and then present *LSH-S* which improves it with random sampling.

### 4.1 Preliminary: LSH Indexing

Let $\mathcal{H}$ be a family of hash functions such that $h \in \mathcal{H} : \mathbb{R}^d \to \mathcal{U}$. Consider a function $h$ that is chosen uniformly at random from $\mathcal{H}$ and a similarity function $sim : \mathbb{R}^d \times \mathbb{R}^d \to [0,1]$. The family $\mathcal{H}$ is called *locality sensitive* if it satisfies the following property [5].

DEFINITION 3. *[Locality Sensitive Hashing] For any vectors $u, v \in \mathbb{R}^d$,*
$$P(h(u) = h(v)) = sim(u,v).$$

That is, the more similar a pair of vectors is, the higher the collision probability is. The LSH scheme works as follows [12, 5]: For an integer $k$, we define a function family $\mathcal{G} = \{g : \mathcal{R}^d \to \mathcal{U}^k\}$ such that $g(v) = (h_1(v), ..., h_k(v))$, where $h_i \in \mathcal{H}$, i.e. $g$ is the concatenation of $k$ LSH functions. For an integer $\ell$, we choose $\ell$ functions $G = \{g_1, ..., g_\ell\}$ from $\mathcal{G}$ independently and uniformly at random. Each $g_i, 1 \le i \le \ell$ effectively constructs a hash table denoted by $D_{g_i}$. A bucket in $D_{g_i}$ stores all $v \in V$ that have the same $g_i$ values. For space, only existing buckets are stored using standard hashing. $G$ defines a collection of $\ell$ tables $I_G = \{D_{g_1}, ..., D_{g_\ell}\}$ and we call it an LSH index. Table 3 in appendix is a summary of notations.

LSH families have been developed for several (dis)similarity measures including Hamming distance, $\ell_p$ distance, Jaccard similarity, and cosine similarity. We rely on the LSH scheme proposed by Charikar [5] that supports cosine similarity. The proposed algorithms can easily support other similarity measures by using an appropriate LSH family.

#### 4.1.1 Extending The LSH Scheme with Bucket Counts

We describe algorithms based on a single LSH table: $g = (h_1, ..., h_k)$ with $k$ hash functions and an LSH table $D_g$. Extensions with multiple LSH tables are in Appendix B.2. Suppose that $D_g$ has $n_g$ buckets; all vectors in the database are hashed into one of the $n_g$ buckets. We denote a bucket by $B_j, 1 \le j \le n_g$. Given a vector $v$, $B(v)$ denotes the bucket to which $v$ belongs. In the LSH table, each bucket $B_j$ stores the set of vectors that are hashed into $B_j$. We extend the conventional LSH table by adding a bucket count $b_j$ for each bucket $B_j$ that is the number of vectors in the database that are hashed into $B_j$. The overhead of adding a bucket count to each bucket is not significant compared to other information such as vectors. Depending on implementation, the count may be readily available.

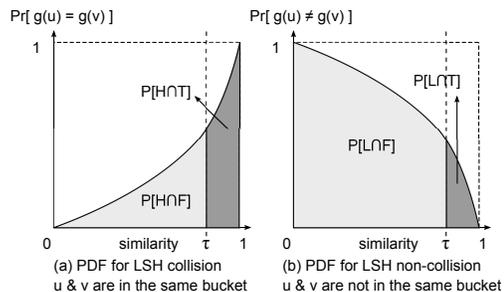

Figure 1: Probability Density Functions of (non-) collision in the LSH scheme

### 4.2 Estimation with Uniformity Assumption

Given a collection of vectors $V$, and a similarity threshold $\tau$, let $M$ denote the number of total pairs in $V$, i.e. $M = \binom{n}{2}$. Consider a random pair $(u,v), u, v \in V, u \ne v$. We denote the event $sim(u,v) \ge \tau$ by $T$, and the event $sim(u,v) < \tau$ by $F$. We call $(u,v)$ a *true* pair (resp. *false* pair) if $sim(u,v) \ge \tau$ (resp. $sim(u,v) < \tau$). Depending on whether $u$ and $v$ are in the same bucket, we denote the event $B(u) = B(v)$ by $H$ and the event $B(u) \ne B(v)$ by $L$. With these notations, we can define various probabilities. For instance, $P(T)$ is the probability of sampling a true pair. $P(H|T)$ is the probability that a true pair is in the same bucket and $P(T|H)$ is the probability that a pair of vectors from a bucket is a true pair. We use $N_\mathcal{E}$ to denote the cardinality of the set of pairs that satisfy the condition of event $\mathcal{E}$. For instance, $N_T$ is the number of true pairs (the VSJ size $J$) and $N_H$ is the number of pairs in the same bucket. $N_H$ can be computed by $N_H = \sum_{j=1}^{n_g} \binom{b_j}{2}$.

The key observation is that any pair of vectors from a bucket is either a true pair or a false pair. Using Bayes' rule [16], we can express this observation as follows: $N_H = N_T \cdot P(H|T) + N_F \cdot P(H|F)$. That is, the total number of pairs of vectors in the same bucket is the sum of the number of true pairs that are in the same bucket ($N_T \cdot P(H|T)$) and the number of false pairs that happened to be in the same bucket ($N_F \cdot P(H|F)$). Since $N_F = M - N_T$, rearranging the terms gives $N_T = (N_H - M \cdot P(H|F))/(P(H|T) - P(H|F))$. Using '^' for an estimated quantity, we have an estimator for the join size $J(= N_T)$ as follows:

$$\hat{N}_T = \hat{J}_U = \frac{N_H - M \cdot \hat{P}(H|F)}{\hat{P}(H|T) - \hat{P}(H|F)}. \quad (1)$$

Note that $M$ and $N_H$ are constants given $V$ and $D_g$. The conditional probabilities in Equation (1) need to be estimated to compute $\hat{N}_T$. We next present our first estimator that relies on an LSH function analysis and a uniformity assumption to estimate the conditional probabilities.

Consider a random pair $(u,v)$ such that $sim(u,v)$ is selected from $[0,1]$ uniformly at random. Recall that when $sim(u,v) = s$, $P(h(u) = h(v)) = s$ from Definition 3. $P(g(u) = g(v)) = s^k$ since $g$ concatenates $k$ hash values. Figure 1(a) shows the collision probability density function (PDF) $f(s) = s^k$ where $s = sim(u,v)$. The vertical dotted line represents the similarity threshold $\tau$. The darker area on the right side of the $\tau$ line is the *good* collision probability, i.e. the probability that $B(u) = B(v)$ ($u$ and $v$ are in the same bucket) and $sim(u,v) \ge \tau$. Thus the area represents $P(H \cap T)$. Likewise, the area on the left side of the $\tau$ line

340

is $P(H \cap F)$, which is the probability that $sim(u,v) < \tau$, but $B(u) = B(v)$. Notice that the area below $f(s)$ does not equal to 1 since it does not cover the entire event space; $u$ and $v$ may not be in the same bucket. Figure 1(b) shows the other case where $u$ and $v$ are in different buckets. Its PDF is $1 - f(s)$ as shown as the curve. $P(L \cap T)$ and $P(L \cap F)$ are defined similarly as shown in the figure.

Given $g$ (and thus $f$) and $\tau$, $P(H \cap F)$, $P(H \cap T)$, $P(L \cap F)$ and $P(L \cap T)$ can be estimated by computing the corresponding areas in Figure 1. Based on these areas, we can estimate the desired conditional probabilities using the following:

$$P(H|T) = \frac{P(H \cap T)}{P(H \cap T) + P(L \cap T)} \quad (2)$$

$$P(H|F) = \frac{P(H \cap F)}{P(H \cap T) + P(L \cap F)}. \quad (3)$$

Plugging $P(H|T)$ and $P(H|F)$ computed as above into Equation (1), we have the following estimator for the VSJ size:

$$\hat{J}_U = \frac{(k+1)N_H - \tau^k \cdot M}{\sum_{i=0}^{k-1} \tau^i}. \quad (4)$$

We give its derivation in Appendix A.1.

We note that $\hat{J}_U$ implicitly assumes that the similarity of pairs is uniformly distributed in $[0, 1]$. However, this distribution is generally highly skewed [14]; most of pairs have low similarity values and only a small number of pairs have high similarity values. We next present $LSH\text{-}S$ that removes the uniformity assumption with sampling.

### 4.3 LSH-S: Removing Uniformity Assumption

We consider two methods to remove the uniformity assumption. First, we estimate the conditional probabilities by random sampling without resorting to the LSH function analysis. For instance, we can estimate $P(H|T)$ by counting the number of pairs in the same bucket among the true pairs in the sample. Second, we weight the conditional probabilities using samples. For example, if all the pairs in the sample have a similarity value of 0.3, we can only consider similarity $s = 0.3$ in Figure 1 without considering the whole area. We call the second method $LSH\text{-}S$ and present only $LSH\text{-}S$ since it outperformed the first one in our experiments. In $LSH\text{-}S$, for each similarity $s$ that appears in the sample, the $f(s)$ value is weighted by the occurrence of $s$ in the sample. For a sample $S$, we use the following weight: $w(s) = |\{(u,v) \in S : sim(u,v) = s\}|/|S|$. For instance, if similarities in $S$ are $\{0.1, 0.1, 0.1, 0.2, 0.2\}$, $w(0.1) = 3/5, w(0.2) = 2/5$ and $w(s) = 0$ for $s \notin \{0.1, 0.2\}$.

Let $S_T$ denote the subset of true pairs in $S$ and $S_F$ denote $S - S_T$. With this weighting scheme, since $f(s) = s^k$,

$$\hat{P}(H|T) = \sum_{(u,v) \in S_T} (sim(u,v))^k / |S_T| \quad (5)$$

$$\hat{P}(H|F) = \sum_{(u,v) \in S_F} (sim(u,v))^k / |S_F|. \quad (6)$$

$LSH\text{-}S$ uses Equations (5) and (6) in Equation (1).

## 5. STRATIFIED SAMPLING USING LSH

A difficulty in $LSH\text{-}S$ is that the conditional probabilities, e.g. $P(H|T)$, need to be estimated and it is hard to acquire reliable estimates of them. In this section, we present an algorithm that overcomes this difficulty by using the LSH index in a slightly different way.

An interesting view of an LSH index is that it partitions all pairs of vectors in $V$ into two strata: pairs that are in the same bucket and pairs that are not. The pairs in the same bucket are likely to be more similar from the property of LSH (Def. 3). Recall that the difficulty of sampling at high thresholds is that the join size is very small and sampling a true pair is hard. Our key intuition is that even at high thresholds it is relatively easy to sample a true pair from the set of pairs that are in the same bucket.

We demonstrate our intuition with a real-world example. Table 1 shows actual probabilities varying $\tau$ in the DBLP data set. We observe that other than at low thresholds, say $0.1 \sim 0.3$, $P(T)$ is close to 0, which implies that naive random sampling is not going to work well with any reasonable sample size. However, note that $P(T|H)$ is consistently higher than 0.04 even at high thresholds. That is, it is not difficult to sample true pairs among the pairs in the same bucket. $P(H|T)$ is large at high thresholds but very small at low thresholds. This means that at high thresholds, a sufficient fraction of true pairs are in the same bucket. But at low thresholds, most of true pairs are not in the same bucket, which implies that the estimate from the pairs in the same bucket is not enough. However, at low thresholds, $P(T|L)$ becomes larger and thus sampling from the pairs that are not in the same bucket becomes feasible.

| $\tau$ | $P(T)$ | $P(T|H)$ | $P(H|T)$ | $P(T|L)$ |
|---|---|---|---|---|
| 0.1 | .082 | 0.31 | 0.00001 | 0.082 |
| 0.3 | .00024 | 0.054 | 0.00041 | 0.00024 |
| 0.5 | .0000034 | 0.049 | 0.0028 | 0.000032 |
| 0.7 | 3.9E-7 | 0.045 | 0.21 | 2.8E-7 |
| 0.9 | 9.1E-8 | 0.040 | 0.86 | 1.3E-8 |

**Table 1: An example probabilities in DBLP**

For sampling in general, it has been observed that "if intelligently used, stratification nearly always results in a smaller variance for the estimated mean or total than is given by a comparable simple random sampling" [8, p99]. We propose below one specific scheme for stratified sampling using LSH, and show its benefit empirically at Section 6.

### 5.1 LSH-SS: Stratified Sampling

We define two strata of pairs of vectors as follows depending on whether two vectors of a pair are in the same bucket.

- *Stratum H* ($S_H$): $\{(u,v) : u,v \in V, B(u) = B(v)\}$
- *Stratum L* ($S_L$): $\{(u,v) : u,v \in V, B(u) \neq B(v)\}$

Note that $S_H$ and $S_L$ are disjoint and thus we can independently estimate the join size from each stratum and add the two estimates. $S_H$ and $S_L$ are fixed given $D_g$. Let $J_H = |\{(u,v) \in S_H : sim(u,v) \geq \tau\}|$, $J_L = |\{(u,v) \in S_L : sim(u,v) \geq \tau\}|$, and let $\hat{J}_H$ and $\hat{J}_L$ be their estimates. We estimate the join size as follows:

$$\hat{J}_{SS} = \hat{J}_H + \hat{J}_L. \quad (7)$$

A straightforward implementation would be to perform uniform random sampling in $S_H$ and $S_L$, and aggregate the two estimates. However, this simple approach may not work well. The problem is that it can be harder to guarantee a small error of $\hat{J}_L$ with the same sample size due to very small $P(T|L)$ at high thresholds [15]. See the next example.

EXAMPLE 1. *Assume that $N_L = 1,000,000$, $J_L = 1$ at $\tau = 0.9$, and the sample size is 10; only one pair out of*

$1,000,000$ *pairs satisfies* $\tau = 0.9$ *and we sample 10 pairs. In most cases, the true pair will not be sampled and* $\hat{J}_L = 0$. *But if the only true pair is sampled,* $\hat{J}_L = 100,000$. *The estimate fluctuates between 0 and* $100,000$ *and is not reliable.*

Our solution for this problem is to use different sampling procedures in the two strata. Recall that similarities of the pairs in $S_H$ are higher and $P(T|H)$ is not too small, even at high thresholds. Thus, for $S_H$, we use uniform random sampling. Relatively small sample size will suffice for reliable estimation in $S_H$. In $S_L$, however, $P(T|L)$ varies significantly depending on the threshold. In general, while at low thresholds $P(T|L)$ is relatively high and the estimate is reliable, $P(T|L)$ becomes very small at high thresholds and the resulting estimate is more likely to have a large error and is much less reliable. For the same sample size, the variance itself decreases as the join size decreases (e.g. Lemma 4.1 of [1]). However, we may need many more samples at high thresholds, where the join size can be very small, to have the same error guarantee [15, 9].

Thus, we use $\hat{J}_L$ only when it is expected to be reliable and discard $\hat{J}_L$ when it is not. Discarding $\hat{J}_L$ at high thresholds generally does not hurt accuracy much since the contribution of $J_L$ in $J$ is not large. In Table 1, when the similarity thresholds is high, $P(H|T)$ is large, which means that a large fraction of true pairs are in $S_H$, not in $S_L$. We use *adaptive sampling* [15] in $S_L$ since it enables us to detect when the estimate is not reliable. A novelty is that we return a safe lower bound when we cannot have a good error guarantee within the allowable sample size.

Algorithm 1 describes the proposed stratified sampling algorithm $LSH$-$SS$. It applies a different sampling subroutine to each stratum. For $S_H$, it runs the random sampling subroutine $SampleH$. For $S_L$, it runs the adaptive sampling subroutine $SampleL$. The final estimate is the sum of estimates from the two Strata as in Equation (7) (line 3).

#### 5.1.1 Sampling in Stratum H

$SampleH$ of Algorithm 1 describes the random sampling in $S_H$. First, a bucket $B_j$ is randomly sampled weighted by the number of pairs in the bucket, $weight(B_j) = \binom{b_j}{2}$ (line 3). We then select a random pair $(u,v)$ from $B_j$ (line 4). The resulting pair $(u,v)$ is a uniform random sample from $S_H$. $SampleH$ has one tunable parameter $m_H$ which is the sample size. We count the number of pairs satisfying the similarity threshold $\tau$ in $m_H$ sample pairs and return the count scaled up by $N_H/m_H$ (line 9).

#### 5.1.2 Sampling in Stratum L

$SampleL$ of Algorithm 1 employs adaptive sampling [15] in $S_L$. It has two tunable parameters $\delta$ and $m_L$. $\delta$ is the answer size threshold, the number of true samples to give a reliable estimate, and $m_L$ is the maximum number of samples. We sample a pair from $S_L$ (line 3) and see if it satisfies the given threshold (line 4). In case LSH computation for bucket checking at line 3 is expensive, we can delay the checking till line 5, which in effect results in slightly more samples. The while-loop at line 2 can terminate in two ways: (1) by acquiring a sufficiently large number of true samples, $n_L \geq \delta$ or (2) by reaching the sample size threshold, $i \geq m_L$, where $i$ is the number of samples taken. In the former case, we return the count scaled up by $N_L/i$ (line 12). Theorem 2.1 and 2.2 of adaptive sampling [15] provide error bounds in such cases. In the latter case, adaptive sampling returns a loose upper bound. We cannot guarantee that the estimate is reliable and simply return the number of true pairs found in the sample, $n_L$, as $\hat{J}_L$ without scaling it up. Since $J_L$ is at least as large as $n_L$, we call $\hat{J}_L = n_L$ a *safe lower bound*.

**Algorithm 1** LSH-SS
**Procedure LSH-SS**
**Input:** similarity threshold $\tau$, sample size for Stratum H $m_H$, answer size threshold $\delta$, max sample size for Stratum L $m_L$
1: $\hat{J}_H = SampleH(\tau, m_H)$
2: $\hat{J}_L = SampleL(\tau, \delta, m_L)$
3: **return** $\hat{J}_{SS} = \hat{J}_H + \hat{J}_L$
**Procedure SampleH**
**Input:** $\tau$, $m_H$ (sample size)
1: $n_H = 0$
2: **for** $i$ from 1 to $m_H$ **do**
3:     sample a bucket $B_j$ with $weight(B_j) = \binom{b_j}{2}$
4:     sample two vectors $u, v \in B_j, u \neq v$
5:     **if** $sim(u,v) \geq \tau$ **then**
6:        $n_H = n_H + 1$
7:     **end if**
8: **end for**
9: **return** $\hat{J}_H = n_H \cdot N_H/m_H$
**Procedure SampleL**
**Input:** $\tau$, $\delta$ (answer size th.), $m_L$ (sample size th.)
1: $i = 0, n_L = 0$
2: **while** $n_L < \delta$ and $i < m_L$ **do**
3:     sample a uniform random pair $(u,v), B(u) \neq B(v)$
4:     **if** $sim(u,v) \geq \tau$ **then**
5:        $n_L = n_L + 1$
6:     **end if**
7:     $i = i + 1$
8: **end while**
9: **if** $i \geq m_L$ **then**
10:     **return** $\hat{J}_L = n_L$ // or $\hat{J}_L = n_L \cdot c_s(N_L/m_L)$ when a dampening factor $0 < c_s \leq 1$ is used.
11: **end if**
12: **return** $\hat{J}_L = n_L \cdot (N_L/i)$

This conservative approach does not degrade accuracy much since the majority of true pairs are in $S_H$ at high thresholds. It is possible, however, that there can be "grey area" of thresholds. While the random sampling does not guarantee its accuracy, there are enough true pairs in $S_L$ to make its impact on underestimation significant. Hence, we consider the use of a dampened scale-up constant $0 < c_s \leq 1$, which is multiplied to the full scaling-up factor at line 10. We analyze the impact of this dampened scale-up factor in the following section.

For the tunable parameters, we used $m_H = n$ and $m_L = n$ and $\delta = \log n$[1] where $n$ is the database size, $n = |V|$. Note that the size is expressed in the number of pairs. This corresponds to sampling $\sqrt{n}$ vectors from two collections of vectors ($n = \sqrt{n} \times \sqrt{n}$) in equi-joins, which is even smaller than the sample size $\sqrt{n} \log n$ in [9]. These parameter values give provably good estimates at both high and low thresholds. We give the details in the following section.

### 5.2 Analysis

Let $\alpha = P(T|H)$ and $\beta = P(T|L)$ for the sake of presentation. In our analysis, a similarity threshold $\tau$ is considered high when $\alpha \geq \log n/n$ and $\beta < 1/n$, and is considered low when $\alpha \geq \log n/n$ and $\beta \geq \log n/n$. Our model is analogous to the classic approach of distinguishing high frequency values from low frequency values to meet the challenge of skewed data, e.g.[9, 7]. Our distinction effectively models different conditional probabilities that are observed in different threshold values as in Table 1.

---
[1]All logarithms used are base-2 logarithms.

We first analyze the high threshold case and then the low threshold case. This distinction is only for the analysis purposes. The behavior of Algorithm 1 is adaptive and users do not need to distinguish two cases by parameters.

### 5.2.1 *Guarantees at high thresholds*

Recall that $\alpha = P(T|H)$ is the probability that a pair of vectors in a bucket is indeed a true pair and $\beta = P(T|L)$ is the probability that a pair of vectors that are not in the same bucket is a true pair. We assume $\alpha \geq \log n/n$ and $\beta < 1/n$ at high thresholds. The condition on $\alpha$ intuitively states that even when the join size is small, the fraction of true pairs in $S_H$ is not too small from the property of LSH. Our assumption on $\alpha$ is not a stringent condition, i.e. $\log n/n$ is usually a very small value and will be easily satisfied by any reasonably working LSH index. The condition on $\beta$ states that it is hard to sample true pairs in $S_L$ at high thresholds.

As a sanity check, consider the example in Table 1. In the data set, $n = 34,000$ and $\beta \sim 0.00003$ ($\sim 1/n$) at $\tau = 0.5$. High thresholds correspond to $[0.5, 1.0]$. It that range $\alpha = P(T|H)$ is consistently higher than 0.04 which is well over the assumed value of $\alpha$ which is 0.00046 ($\sim \log n/n$) for the data set. $\beta = P(T|L)$ is also below or very close to the calculated 0.00003 in the range. Table 2 in Appendix C shows more examples of $\alpha$ and $\beta$ values in other data sets, which again satisfy our assumptions.

The following theorem states that $LSH$-$SS$ gives a good estimate at high thresholds. See Appendix A.2 for a proof.

THEOREM 1. *Let $0 < \epsilon < 1$ be an arbitrary constant. Assume $\alpha \geq \log n/n$ and $\beta < 1/n$. Then for sufficiently large $n$ with $c = 1/(\log e \cdot \epsilon^2)$, $m_H = cn$ and $m_L = cn$,*

$$Pr(|\hat{J}_{SS} - J| \leq (1+\epsilon)J) \geq 1 - \frac{2}{n}.$$

The term 1 in the error bound is from the conservative nature of the *SampleL* procedure of not scaling up the result when the accuracy is not guaranteed. However, since the majority of true pairs are in $S_H$ at high thresholds, its impact is rather insignificant. For overestimations, the bound is very conservative and our experiments in Section 6 show much better results than the theoretical guarantees in Theorem 1. For underestimations, the error bound in Theorem 1 is not meaningful because underestimation is capped by $-100\%$. The following theorem, which is based on the dampened scale-up factor, has the advantage of providing a meaningful bound for underestimations as well.

In the approach using a dampened scale-up constant $0 < c_s \leq 1$, we multiply the full scale-up factor by $c_s$. The following theorem quantifies the effect of using the dampened scale-up factor. See Appendix A.3 for a proof.

THEOREM 2. *For a constant $0 < \epsilon < 1$, using a dampened scale-up factor of $0 < c_s \leq 1$ at line 10 of SampleL in Algorithm 1 guarantees*

$$Pr(|\hat{J}_L - J_L| \geq \epsilon' J_L) \leq \epsilon^{-2} \cdot \frac{1-\beta}{m_L \beta}.$$

*where $\epsilon' = 1 - (1-\epsilon)c_s$. When $\hat{J} > J$, i.e. overestimation, we have a tighter bound $\epsilon' = c_s(1+\epsilon) - 1$.*

Compared with Theorem 1, the guarantee of Theorem 2 is weaker in terms of probability, but it provides a tighter bound; $1 - c_s < \epsilon' < 1$. The dampened scale-up factor $c_s$ represents the trade-offs between the accuracy and the probabilistic guarantee. Using a smaller $c_s$ enables a stronger guarantee in probability reducing variance by a factor of $c_s$ (see Appendix A.3), but the relative error bound increases. We discuss the choice of $c_s$ in Section 6.

### 5.2.2 *Guarantees at low thresholds*

At low thresholds, we assume that $\alpha \geq \log n/n$ and $\beta \geq \log n/n$. The rationale is that as the actual join size increases, more true pairs are in $S_L$ and sampling true pairs in $S_L$ becomes not so difficult any more. Again these conditions are usually met when the threshold is low as in the example in Table 1. In fact, the contribution from $S_L$, $J_L$ dominates the join size at low thresholds. The following theorem states that $LSH$-$SS$ gives a reliable estimate even when the threshold is low. See Appendix A.4 for a proof.

THEOREM 3. *Let $0 < \epsilon < 1$ be an arbitrary constant. Assume $\alpha \geq \log n/n$ and $\beta \geq \log n/n$. Then with $c = 4/(\log e \cdot \epsilon^2)$, $c' = max(c, 1/(1-\epsilon))$, $m_H = cn$ and $m_L = c'n$,*

$$Pr(|\hat{J}_{SS} - J| \leq \epsilon J) \geq 1 - \frac{3}{n}.$$

We demonstrate that our guarantees indeed hold and thus $LSH$-$SS$ provides reliable estimates at both high and low thresholds with real-world data sets in the following section.

## 6. EXPERIMENTAL EVALUATION

### 6.1 Set Up

**Data sets:** We have conducted experiments using three data sets. The DBLP data set consists of about 800K vectors. The NYT data set is NYTimes news articles and consists of about 150K vectors. The PUBMED data set is PubMed abstracts and consists of 400K vectors. We give detailed data set descriptions in Appendix C.1. We only report results of the DBLP and NYT data set; results on the PUBMED data set is in Appendix C.4.

**Algorithms compared:** We implemented the following algorithms for the VSJ problem.

- $LC(\xi)$ is the Lattice Counting algorithm in [14] with a minimum support threshold $\xi$.
- $LSH$-$S$ is the LSH index based algorithm in Section 4.
- $LSH$-$SS$ is the LSH index based stratified sampling algorithm in Section 5. We used $m_1 = n$, $\delta = \log n$, $m_2 = n$ and $k = 20$ unless specified otherwise. $LSH$-$SS(D)$ is $LSH$-$SS$ with a dampened scale-up factor. We used $c_s = n_L/\delta$. See Appendix C.3 for a discussion on alternative choices.
- We consider two random sampling methods: $RS(pop)$ and $RS(cross)$. In $RS(pop)$, we sample pairs from the cross product. $RS(cross)$ is cross sampling [10], where we sample $\lceil \sqrt{m_R} \rceil$ records and compare all the pairs in the sample. $m_R = d \cdot n$ where $d$ is a constant to compare algorithms with roughly the same runtime.

**Evaluation metric:** We use average relative error to show the accuracy. A relative error is defined as $est\_size$ / $true\_size$. We show errors of overestimations and underestimations separately to clearly depict the characteristics of each algorithm. To measure reliability, we report the standard deviation (STD) of estimates. We report figures over 100 experiments. For efficiency, we measure the runtime, which is time taken for estimation. For all algorithms, we loaded necessary data structures (data or index) in memory.

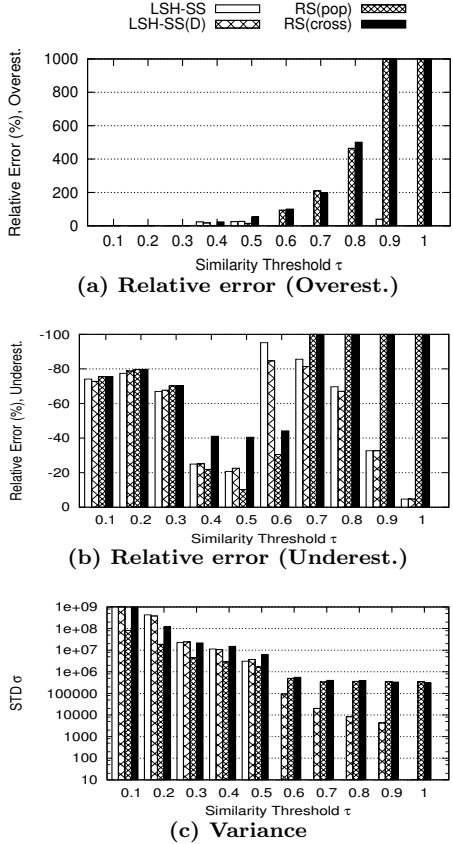

Figure 2: Accuracy/Variance on DBLP

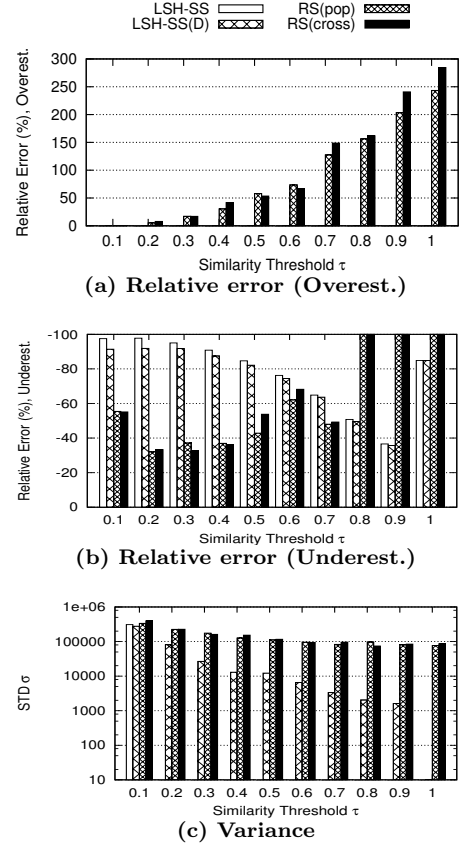

Figure 3: Accuracy/Variance on NYT

| $\tau$ | 0.1 | 0.3 | 0.5 | 0.7 | 0.9 |
|---|---|---|---|---|---|
| $J$ | 105B | 267M | 11M | 103K | 42K |
| selectivity | 33% | 0.085% | 0.0036% | 0.000064% | 0.000013% |

We implemented all the algorithms in Java. We ran all our experiments on a server running 64 GNU/Linux 2.6.27 over 2.93 GHz Intel Xeon CPUs with 64GB of main memory.

### 6.2 Accuracy, Variance and Efficiency

**DBLP:** We first report the results on the accuracy, variance and runtime using the DBLP data set. Figure 2(a) (resp. Figure 2(b)) shows relative errors of overestimations (resp. underestimations) over the similarity threshold range. Figure 2(c) shows the variance of the estimates.

$LSH\text{-}SS$ delivers accurate estimations over the whole threshold range. In Figure 2(a), we see that $LSH\text{-}SS$ hardly overestimates, which is expected from the discussions in Section 5.2. $LSH\text{-}SS(D)$ occasionally overestimates, but its error is smaller than 30%. Figure 2(b) shows the underestimation tendency of $LSH\text{-}SS$, but it is much less severe than those of $RS(pop)$ and $RS(cross)$. We observe that $LSH\text{-}SS(D)$ shows smaller underestimations.

The following table shows the actual join size $J$ and its selectivity at various similarity threshold $\tau$.

Note the dramatic differences in the join size depending on $\tau$. At $\tau = 0.1$ there are more than 100 billion true pairs, and its selectivity is about 30 %. But at $\tau = 0.9$ there are only 42,000 true pairs, and the selectivity is only 0.00001%. Yet $LSH\text{-}SS$ is able to estimate the join size quite accurately and reliably exploiting the LSH index. Moreover, in Figure 2 (c), $LSH\text{-}SS$ shows very small variance at high thresholds.

$RS(pop)$ and $RS(cross)$ are as accurate as $LSH\text{-}SS$ at low thresholds. However as the threshold increases, their errors rapidly increase; their estimations fluctuate between huge overestimation and extreme underestimation (i.e. $-100\%$) in Figure 2(a) and (b). As a consequence, their variance is quite large in Figure 2(c).

$LSH\text{-}S$ and $LC$ show consistently outperformed by others, and we omit their results. $LSH\text{-}S$ has large errors at high thresholds, e.g. $\tau \geq 0.6$. This is because the estimations of conditional probabilities are not reliable due to insufficient number of true pairs sampled. $LC$ underestimates over the whole threshold range. We hypothesize that it is because of the characteristic of the binary LSH function for cosine similarity. Intuitively the binary LSH functions need more hash functions (larger $k$) to distinguish objects, which has negative impact on the runtime. It appears that $LC$ is not adequate for binary LSH functions.

For runtime, $LSH\text{-}SS$ and $LSH\text{-}SS(D)$ took less than 750 msec and $RS(pop)$ and $RS(cross)$ took about 780 sec on the average. The runtime of $LSH\text{-}S$ was 1028 msec, and that of $LC$ was about 3 sec.

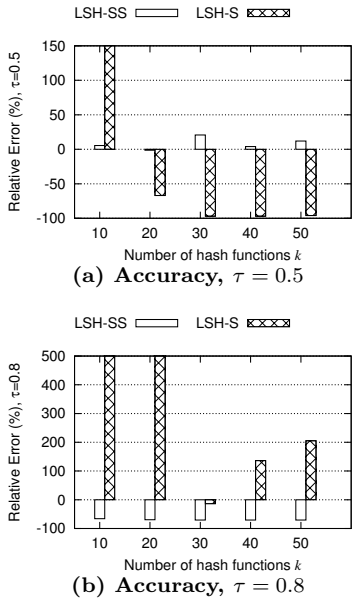

**Figure 4: Impact of $k$ on DBLP**

**NYT:** Figure 3 shows relative errors and variance on the NYT data set. *LSH-SS* gives good estimates at high thresholds. It shows underestimation at $\tau \leq 0.5$. In general, however, this is not the most interesting similarity range and this is probably not a serious flaw. The problem can be addressed either by using a bigger dampened scale-up factor $c_s$ or increasing the sample size. We see that $LSH\text{-}SS(D)$ has smaller underestimation problems. Again, we observe that estimations of $RS(pop)$ and $RS(cross)$ fluctuate at high thresholds. Their variance is larger than the variance of $LSH\text{-}SS$ or $LSH\text{-}SS(D)$ throughout the threshold range.

For runtime, *LSH-SS* took 1091 msec and *RS* took 920 msec on the average.

## 6.3 Impact of Parameters

We assume a pre-built LSH index with parameters optimized for its similarity search. The relevant parameter for our estimation purposes is $k$ which specifies the number of hash functions used for an LSH table. We analyze the impact of $k$ on accuracy and variance.

Figure 4 shows accuracy at $\tau = 0.5, 0.8$. The conclusion is the same at other thresholds. We observe that *LSH-SS* is not much affected by $k$. This is because an LSH table provides sufficient *distinguishing* power with relatively small $k$. *LSH-SS* will work with any reasonable choice of $k$. *LSH-S* is highly sensitive to $k$ for the reason specified in Section 6.2. The same observation is made in variance as well.

|  | $k=10$ | $k=20$ | $k=30$ | $k=40$ | $k=50$ |
|---|---|---|---|---|---|
| size (MB) | 3.2 | 7.5 | 12.6 | 14.1 | 16.5 |

The above table shows space occupied by an LSH table on the DBLP data set ignoring implementation-dependent overheads. When $k=20$, there are about $480K$ non-empty buckets which add $7.5M$ of space for the $g$ function values, bucket counts, and vector ids. The DBLP binary is $50M$.

If we are given the freedom of choosing $k$, we observe that slightly smaller $k$ values, say between 5 and 15, generally give better accuracy. See Appendix B.1 for more discussions.

## 7. CONCLUSION

We propose size estimation techniques for vector similarity joins. The proposed methods rely on the ubiquitous LSH indexing and enable reliable estimates even at high similarity thresholds. We show that *LSH-SS* gives good estimates throughout the threshold range with probabilistic guarantees. The proposed techniques only need minimal addition to the existing LSH index and can be easily applied.

# APPENDIX

## A. PROOFS

### A.1 Equation (4)

Since $f(s) = s^k$, the four areas defined in Figure 1 can be calculated as follows:

$$
\begin{aligned}
P[H \cap F] &= \int_0^\tau f(s)ds = \frac{\tau^{k+1}}{k+1} \\
P[H \cap T] &= \int_\tau^1 f(s)ds = \frac{1 - \tau^{k+1}}{k+1} \\
P[L \cap F] &= \int_0^\tau 1 - f(s)ds = \tau - \frac{\tau^{k+1}}{k+1} \\
P[L \cap T] &= \int_\tau^1 1 - f(s)ds = 1 - \tau - \frac{1 - \tau^{k+1}}{k+1}.
\end{aligned}
$$

Using the above probabilities in Equation (2) and (3) gives,

$$
P(H|T) = \frac{1}{1-\tau} \int_\tau^1 f(s)ds = \frac{\sum_{i=0}^k \tau^i}{k+1} \quad (8)
$$

$$
P(H|F) = \frac{1}{\tau} \int_0^\tau f(s)ds = \frac{\tau^k}{k+1}. \quad (9)
$$

We have the following estimator $\hat{J}_U$ by using above $P(H|T)$ and $P(H|F)$ in Equation (1):

$$
\begin{aligned}
\hat{J}_U &= \frac{N_H - M \cdot \hat{P}(H|F)}{\hat{P}(H|T) - \hat{P}(H|F)} \\
&= \frac{N_H - M \cdot \frac{\tau^k}{k+1}}{\frac{\sum_{i=0}^k \tau^i}{k+1} - \frac{\tau^k}{k+1}} \\
&= \frac{(k+1)N_H - M \cdot \tau^k}{\sum_{i=0}^{k-1} \tau^i}.
\end{aligned}
$$

### A.2 Proof of Theorem 1

We analyze the behavior of *SampleH* and *SampleL* in Algorithm 1 separately by the two lemmas below. We then combine the two results using Equation (7).

First, the following lemma shows that *SampleH* in Algorithm 1 gives a reliable estimate at high thresholds.

LEMMA 1. *Let $0 < \epsilon < 1$ be an arbitrary constant. Assume $\alpha \geq \log n / n$. Then with $c = 4/(\log e \cdot \epsilon^2)$ and $m_H = cn$, we have*

$$
Pr(|\hat{J}_H - J_H| \leq \epsilon J_H) \geq 1 - \frac{1}{n}.
$$

PROOF. Let $X$ be a random variable denoting the number of pairs in the sample that satisfy $\tau$ in *SampleH* of Algorithm 1. Then $X$ is a binomial random variable with parameters $(m_H, \alpha)$ [16]. Since $m_H = cn$ and $\alpha \geq \log n / n$,

$$
E(X) \geq c \log n.
$$

For an arbitrary constant $0 < \epsilon < 1$, by Chernoff bounds [16]

$$
Pr(|X - E(X)| > \epsilon E(X)) \leq e^{-\frac{c \log n \epsilon^2}{4}}.
$$

Letting $c = 4/(\log e \cdot \epsilon^2)$ gives

$$
Pr(|X - E(X)| > \epsilon E(X)) \leq \frac{1}{n}.
$$

$E(X) = J_H \cdot \frac{m_H}{N_H}$. Thus, $\frac{N_H}{m_H} \cdot E(X) = \frac{N_H}{m_H} \cdot m_H \cdot \frac{J_H}{N_H} = J_H$. Therefore,

$$
Pr(|\frac{N_H}{m_H} \cdot X - J_H| > \epsilon J_H) \leq \frac{1}{n}.
$$

$\hat{J}_H = X \cdot N_H / m_H$ in *SampleH* of Algorithm 1. Plugging in $\hat{J}_H$ in the above inequality completes the proof. □

Second, the following lemma states that *SampleL* returns a safe lower bound with high probability.

LEMMA 2. *Assume $\beta < 1/n$. Then for sufficiently large $n$, an arbitrary constant $c'$ and $m_L = c'n$, we have*

$$
Pr(\hat{J}_L \leq c') \geq 1 - \frac{1}{n}.
$$

PROOF. Let $Y$ be a random variable denoting the number of pairs in the samples satisfying $\tau$ in *SampleL* of Algorithm 1. We show that $Y$ is not likely to be bigger than $\delta = \log n$. $Y$ is a binomial random variable with parameters $(m_L, \beta)$.

For an arbitrary constant $\epsilon' \geq 2e - 1$, by Chernoff bounds [16]

$$
Pr(Y > (1 + \epsilon')E(Y)) \leq 2^{-E(Y)(1+\epsilon')}.
$$

Therefore, the probability that the loop of *SampleL* terminates by acquiring enough number of true pairs ($\delta = \log n$) is as follows:

$$
Pr(Y > \delta = \log n) \leq 2^{-\log n} = \frac{1}{n}.
$$

Since $m_L = c'n$ and $\beta < 1/n$,

$$
E(Y) < c'.
$$

If $Y \leq \delta$, *SampleL* of Algorithm 1 returns $\hat{J}_L = Y$ without scaling it up. Therefore,

$$
Pr(\hat{J}_L = E(Y) < c') \geq 1 - \frac{1}{n}.
$$

□

Finally, we prove Theorem 1 using Lemma 1 and Lemma 2. For sufficiently large $n$, $c' \leq J_L$. Thus from Lemma 2,

$$
Pr(|\hat{J}_L - J_L| \leq J_L) \geq 1 - \frac{1}{n}.
$$

Since $\hat{J}_{SS} = \hat{J}_H + \hat{J}_L$ and $J = J_H + H_L$, combining the above inequality with Lemma 1 proves the theorem.

### A.3 Proof of Theorem 2

Let $X$ be a random variable for the number of true pairs in the sample, i.e. $n_L$, in *SampleL* of Algorithm 1. $X$ follows a binomial distribution $B(m_L, \beta)$, and $\mu_X = m_L \beta$ and $\sigma_X = \sqrt{m_L \beta (1-\beta)}$. Suppose $0 < c_s \leq 1$ is used as a dampening factor. Let $s$ be the sampling ratio, $s = m_L/N_L$, and $Y$ be the estimate using $c_s$. $Y = c_s/s \cdot X$, and $\mu_Y = c_s/s \cdot m_L \beta = c_s N_L \beta = c_s J_L$ and $\sigma_Y = c_s/s \cdot \sqrt{m_L \beta (1-\beta)}$. Applying Chebyshev's inequality [16], for any $k \in \mathbb{R}^+$,

$$
Pr\left(|Y - \mu_Y| \geq k \cdot \sigma_Y\right) \leq \frac{1}{k^2}.
$$

$\sigma_Y/\mu_Y = \sqrt{(1-\beta)/m_L \beta}$. Letting $\epsilon = k\sqrt{(1-\beta)/m_L \beta}$ gives

$$
Pr\left(\frac{|Y - \mu_Y|}{\mu_Y} \geq \epsilon\right) \leq \epsilon^{-2} \cdot \frac{1-\beta}{m_L \beta}.
$$

The lower bound of the estimation is $J_L - (1-\epsilon)\mu_Y$. In overestimation, $c_s(1+\epsilon)J_L \geq J_L$ since otherwise it would be underestimation. Thus, the upper bound of the estimation is $(c_s(1+\epsilon)-1)J_L$ because $c_s \leq 1$, $1-(1-\epsilon)c_s \geq c_s(1+\epsilon)-1$. Letting $\epsilon' = 1 - (1-\epsilon)c_s$ proves the theorem.

## A.4 Proof of Theorem 3

As we have the same conditions on $\alpha$, $m_H$, and $c$, Lemma 1 still holds in the low threshold range as well. However, due to the increased $\beta$, a different analysis needs to be done for *SampleL* (Lemma 2). We first show that *SampleL* returns a scaled-up estimate not a safe lower bound with high probability, and then show that the scaled-up estimate is reliable.

Similarly as in Lemma 2, $Y$ is a binomial random variable with parameters $(m_L, \beta)$. Since $m_L = c'n$ and $\beta \geq \log n/n$,

$$E(Y) \geq c' \log n.$$

From the given condition, $c' \geq 2/(\log e \cdot \epsilon^2)$ and $c' \geq 1/(1-\epsilon)$. Since $(1-\epsilon)E(Y) \geq (1-\epsilon)c' \log n \geq \log n = \delta$, by Chernoff bounds,

$$Pr(Y \geq \delta) \geq 1 - \frac{1}{n}.$$

This means that the while-loop (line 2) of *SampleL* Algorithm 1 terminates by reaching the desired answer size threshold $\delta$ with high probability. Then *SampleL* returns $\hat{J}_L = Y \cdot J_L/m_L$. Moreover, since $c' \geq 4/(\log e \cdot \epsilon^2)$, by Chernoff bounds,

$$Pr(|Y - E(Y)| \geq \epsilon E(Y)) \leq \frac{2}{n}.$$

Therefore,

$$Pr(|\hat{J}_L - J_L| \leq \epsilon J_L) \geq 1 - \frac{2}{n}.$$

Since $\hat{J}_{SS} = \hat{J}_H + \hat{J}_L$ and $J = J_H + H_L$, the above inequality along with Lemma 1 completes the proof.

## B. ADDITIONAL DISCUSSION

### B.1 The Optimal-$k$ for The VSJ Problem

Recall that $k$ is the number of LSH functions for $g$. Ideally, we want high $P(T|H)$ and $P(H|T)$ because the estimate from $S_H$ is more reliable and having more true pairs in $S_H$ reduces the dependency on $S_L$. $P(T|H)$ (resp. $P(H|T)$) is analogous to precision (resp. recall) in information retrieval. We note that $k$ value has the following trade-offs between $P(T|H)$ and $P(H|T)$.

- A larger $k$ increases $P(T|H)$ but decreases $P(H|T)$. With a sufficiently large $k$, only exactly the same vectors will be in the same bucket. $P(T|H) = 1$ in this case. However, only a very small fraction of true pairs is in the same bucket resulting in a very small $P(H|T)$.

- A smaller $k$ decreases $P(T|H)$ but increases $P(H|T)$. In an extreme case of $k = 0$, $S_H$ consists of all the pairs of vectors in $V$, and thus $P(H|T) = 1$. However, $P(T|H) = P(T)$ and the LSH scheme does not offer any benefit.

Observe that $P(T|H) \geq P(T|L)$ from the property of the LSH indexing. The intuition on choosing $k$ is that we want to increase $J_H$ as long as we have good estimates with high probability. Since decreasing $k$ increases $P(H|T)$ and $J_H$, this means that we can decrease $k$ as long as $P(T|H)$ is not too small. Decreasing $k$ also reduces the LSH function computation time. With this intuition, we can formalize the problem of choosing $k$ as follows:

DEFINITION 4 (THE OPTIMAL-$k$ PROBLEM). *Given a desired error bound $\epsilon > 0$ and the bound on the probabilistic guarantee $p$, find the minimum $k$ such that $P(T|H) \geq \rho$, where $\rho = \rho(\epsilon, p)$.*

If can we assume a similarity distribution of the database, we can analytically find the optimal $k$. However, $P(T|H)$ is dependent on data and the LSH scheme used, and so is the optimal value of $k$.

### B.2 Extensions

#### B.2.1 Using Multiple LSH Tables

The proposed algorithms so far assume only a single LSH table, but a typical LSH index consists of multiple LSH tables. In this section, we describe how we can utilize more than one LSH tables for the estimation purposes. We consider two estimation algorithms using multiple LSH tables: median estimator and virtual bucket estimator.

**Median estimator.** The median estimator applies *LSH-SS* to each LSH table independently without modifying *LSH-SS* and merges the estimates. Suppose an LSH index $I_G = \{D_{g_1}, \ldots, D_{g_\ell}\}$ with $\ell$ tables. From each table $D_{g_i}, 1 \leq i \leq \ell$, we generate an estimate $\hat{J}_i$ with a sample of $n$ pairs. Its estimate, $\hat{J}_m$, is the median of the estimates, $\hat{J}_i, 1 \leq i \leq \ell$. This approach makes the algorithm more reliable reducing the probability that $\hat{J}_m$ deviates much from $J$. From Theorems 1 and 3, the probability that $\hat{J}_i$ differs from $J$ by more than a factor of $1+\epsilon$ is less than $2/n$ with the assumed join size and sample size. When taking the median, the probability that more than $\ell/2$ $\hat{J}_i$'s deviate by more than $(1+\epsilon)J$ is at most $2^{-\ell/2}$ by the standard estimate of Chernoff [16]. This states that $\hat{J}_m$ is within the same factor of error with higher probability than the guarantees in Theorems 1 and 3. The effective sample size increases by a factor of $\ell$. When a sample size that is greater than $n$ is affordable, exploiting multiple LSH tables can make the estimate more reliable. However, dividing a total sample size of $n$ into multiple estimates can impair the accuracy of individual estimates.

**Virtual bucket estimator.** We consider virtual buckets formed by multiple LSH tables. A pair $(u, v)$ is regarded as in the same bucket if $u$ and $v$ are in the same bucket in any of $\ell$ LSH tables. This can improve the accuracy when an existing LSH scheme has a relatively large $k$ than necessary. Recall from the discussions in the previous section that when $k$ is too large, $g$ becomes overly selective and $S_H$ can be too small. Then the true pairs from $S_H$ can be only a small portion of the true pairs. Considering virtual buckets can address this problem by relaxing the bucket conditions; $B(u) = B(v)$ if and only if there exist a $D_{g_i}$ such that $u$ and $v$ belong to the same bucket $B \in D_{g_i}$. With virtual buckets, when we check $B(u) = B(v)$ (or $\neq$) for $(u, v)$, we need to check up to $\ell$ tables. At lines 3 and 4 of *SampleH* in Algorithm 1, a pair $(u, v)$ is chosen from $V$ uniformly at random and is discarded if $u$ and $v$ are not in the same bucket in any $D_{g_i}, 1 \leq i \leq \ell$. At line 3 of *SampleL*, $B_i(u) \neq B_i(v)$ is checked for all $D_{g_i}$ and if $B_i(u) = B_i(v)$ in any $D_{g_i}$, $(u, v)$ is discarded. The analysis remains the same but the set of pairs in same buckets, $S_H$ becomes effectively larger, which gives potentially better accuracy. The runtime increases because of the bucket checking in multiple LSH tables.

### B.2.2 Non-self Joins

In this section, we discuss how to extend the proposed algorithms to handle joins between two collections of vectors $U$ and $V$. The basic ideas remain the same but we need to make sure that a pair under consideration consists of one vector from each collection.

DEFINITION 5 (THE GENERAL VSJ PROBLEM). *Given two collection of real-valued vectors $V = \{v_1, ..., v_{n_1}\}$, $U = \{u_1, ..., u_{n_2}\}$ and a threshold $\tau$ on a similarity measure sim, estimate the number of pairs $J = |\{(u,v) : u \in U, v \in V, sim(u,v) \geq \tau\}|$.*

Suppose that we have two LSH tables $D_g$ and $E_g$ that are built on $U$ and $V$ respectively using $g = (g_1, \ldots, g_\ell)$. We describe how we modify *LSH-S* and *LSH-SS*.

**LSH-S.** We make two changes for *LSH-S*: $N_H$ computation and sampling. In self-joins, $N_H$ in Equation (1) is the number of pairs in the same bucket: $N_H = \sum_{j=1}^{n_g} \binom{b_j}{2}$. In general joins, $N_H = \sum_{j=1}^{n_g} b_j \cdot c_i$ such that $B_j \in D_g, C_i \in E_g$, $b_j$ is the bucket count of $B_j$, $c_i$ is the bucket count of $C_i$, and $g(B_j) = g(C_i)$, where $g(B)$ denotes the $g$ value that identifies bucket $B$. For $B_i$, if there does not exist a bucket $C_i \in E_g$ such that $g(C_i) = g(B_j)$, $c_i = 0$. Next, a pair $(u,v)$ is sampled uniformly at random such that $u \in U$ and $v \in V$.

**LSH-SS.** $S_H$ is the set of pairs $(u,v), u \in U, v \in V$ such that $g(u) = g(v)$. That is, the buckets of $u$ and $v$ have the same $g$ value: $g(B_j) = g(C_i)$ where $u \in B_j$ in $D_g$ and $u \in C_i$ in $E_g$. $S_L$ is the set of pairs $(u,v), u \in U, v \in V$ such that $g(u) \neq g(v)$. To sample a pair from $S_H$, we randomly sample a bucket $B_j$ with $weight(B_j) = b_j \cdot c_i$ where $g(C_i) = g(B_j), B_j \in D_g$ and $C_i \in E_g$. We sample $u$ from $B_j$ and $v$ from $C_i$ uniformly at random within the buckets. The resulting pair $(u,v)$ is a uniform random sample from $S_H$. Lines 3 and 4 of *SampleH* in Algorithm 1 need to be changed accordingly. To sample from $S_L$, we sample $u \in U$ and $v \in V$ uniformly at random, and $(u,v)$ is discarded if $g(u) = g(v)$. Line 3 of *SampleH* needs corresponding changes.

## C. SUPPLEMENTARY EXPERIMENTS

### C.1 Data Set Description

The DBLP data set consists of 794,016 publications is the same as the data set used in [3]. Each vector is constructed from authors and title of a publication. There are about 56,000 distinct words in total and each word is associated with a dimension in a vector. The vector of a publication represents whether the corresponding word is present in the publication. Thus this data set is a binary vector data set. The average number of features is 14, and the smallest is 3 and the biggest is 219. The NYT data set is NYTimes news articles downloaded from UCI Machine Learning Repository[2] and consists of 149,649 vectors. Again each dimension represents a word and has the corresponding TF-IDF weight. The dimensionality is about 100k and the average number of feature is 232. The PUBMED data set is constructed from PubMed abstracts and is also downloaded from the UCI repository. It consists of 400,151 TF-IDF vectors. The dimensionality is about 140k. It takes 4.7, 4.6, and 5.6 seconds to build LSH indexes from respective in-memory raw data.

[2]http://archive.ics.uci.edu/ml/datasets.html

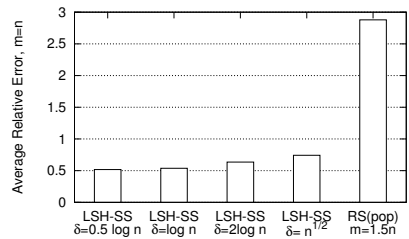

**Figure 5: Relative error varying $\delta$ (the answer size threshold) in *SampleL***

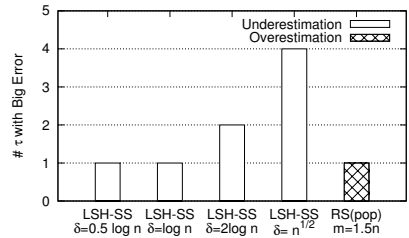

**Figure 6: The number of $\tau$ values with $\hat{J}/J \geq 10$ (overest.) or $J/\hat{J} \geq 10$ (underest.) varying $\delta$**

### C.2 Impact of Parameters: $\delta$ and $m$

In this section, we study the impact of parameters related with the sample size using the DBLP data set. The goal is to see if our analysis and choice of parameter values are appropriate. Recall that *LSH-SS* and *RS* have the following parameters; $m_H$ is the sample size in *SampleH*, $\delta$ is the answer size threshold in *SampleL*, $m_L$ is the (maximum) sample size in *SampleL*, and $m_R$ is the sample size of random sampling. In our analysis and experiments, we used the following parameter values: $m_H = m_L = n$, $\delta = \log n$, and $m_R = 1.5n$. Since $m_H, m_L$ and $m_R$ control the overall sample size and $\delta$ specifies the answer size threshold, we use two functions $f_1$ and $f_2$ to control the parameters: $m_H = m_L = f_1(n)$, $m_R = 1.5 f_1(n)$ and $\delta = f_2(n)$. We test the following alternatives:

- $f_1$: $\sqrt{n}$, $n/\log n$, $0.5n$, $\underline{n}$, $2n$, and $n \log n$
- $f_2$: $0.5 \log n$, $\underline{\log n}$, $2 \log n$, and $\sqrt{n}$

We perform two types of combinations of $f_1$ and $f_2$. First, we fix $f_1(n) = n$ and vary $f_2$. Next, we fix $f_2(n) = \log n$ and vary $f_1$. For each combination we show two results: the average absolute error for $\tau = \{0.1, 0.2, ..., 1.0\}$ and the number of $\tau$ values with large errors among the 10 $\tau$ values. We define an error to be a big overestimation when $\hat{J}/J \geq 10$ and to be a big underestimation when $J/\hat{J} \geq 10$.

#### C.2.1 Answer Size Threshold $\delta$

Figure 5 gives the average (absolute) relative error varying $\delta$ and Figure 6 gives the number of $\tau$ values that give large errors. $m(= f_1)$ is fixed at $n$. $\delta > 2 \log n$ has a big underestimation problem. A large $\delta$ may prevent even a reliable estimate from being scaled up and result in a huge

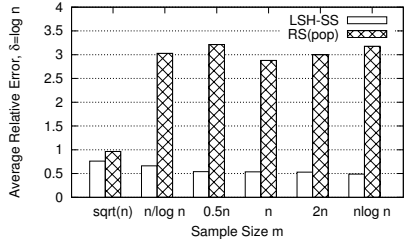

**Figure 7: Relative error varying the sample size $m$**

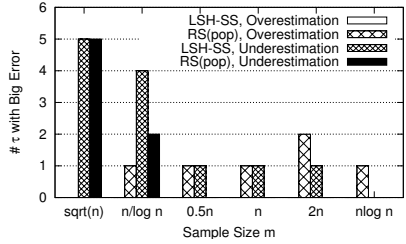

**Figure 8: The number of $\tau$ with $\hat{J}/J \geq 10$ (overest.) or $J/\hat{J} \geq 10$ (underest.) varying the sample size $m$. The total number of $\tau$ is 10: $\{0.1, 0.2, ..., 1.0\}$**

loss in contributions from $S_L$. For instance, $\delta = \sqrt{n}$ is too conservative and its estimate is less than 10% of the true size at 4 out of 10 $\tau$ values.

Simple heuristics such as using different $\delta$ depending on the threshold can improve the performance. For instance using $0.1 \log n \leq \delta \leq 0.5 \log n$ at high thresholds, e.g. $\tau \geq 0.7$, greatly improved the runtime without sacrificing accuracy in our experiments. At low thresholds, e.g. $\tau \leq 0.3$, using a slightly bigger value of $\delta$, e.g. $2 \log n$, resulted in better accuracy and variance without increasing the runtime much.

### C.2.2 Sample Size Threshold $m$

Figure 7 gives the average absolute relative error varying the sample size and Figure 8 gives the number of $\tau$ values that give large errors. $\delta$ is fixed at $\log n$. $f_1 < 0.5n$ causes serious underestimations in both algorithms. With $f_1 = n \log n$, *LSH-SS* does not give any large errors, but the runtime roughly increases by $\log n$.

## C.3 Impact of Parameter: $c_s$

Using a larger dampened scale-up factor $c_s$ has less underestimation. In the DBLP data set, setting $c_s = 0.5$ reduces underestimation errors from $-95\%$ (*LSH-SS*) to $-65\%$ at $\tau = 0.6$. However, it can cause overestimation problems with large variance as discussed in Section 5.2. $c_s = 1$ has overestimation errors between 100% and 900% at high thresholds. $c_s = 0.5$ gives errors between 16% and 437%, $c_s = 0.1$ gives errors less than 62%. The choice of $c_s$ depends on specific application requirements, but if variance is not a concern, $0.1 \leq c_s \leq 0.5$ can be recommended.

## C.4 PUBMED: Accuracy, Variance, Efficiency

Figure 9 shows the accuracy and variance on the PUBMED data set with $k = 5$. The average error of *LSH-SS* is 73% and that of *RS* is 117%. *LSH-SS* shows underestimation tendency but its STD is more than an order of magnitude smaller than that of *RS*. When the data set is largely dis-similar, smaller $k$ improves accuracy. In such cases, constructing an LSH table on-the-fly can be a viable option.

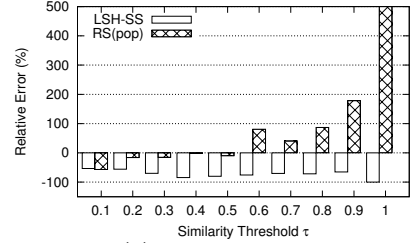

(a) Relative error

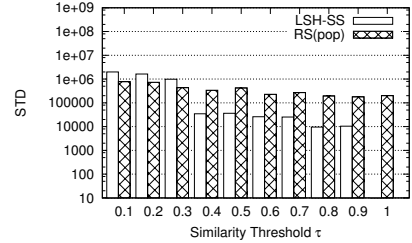

(b) Variance

**Figure 9: Accuracy/Variance on PUBMED**

| | NYT | | PUBMED | |
|---|---|---|---|---|
| $\tau$ | $\alpha$ | $\beta$ | $\alpha$ | $\beta$ |
| 0.1 | .710 | 1.85E-4 | .0179 | .00593 |
| 0.3 | .710 | 3.26E-5 | 2.15E-4 | 6.37E-7 |
| 0.5 | .708 | 9.93E-6 | 2.15E-4 | 1.40E-8 |
| 0.7 | .705 | 3.25E-6 | 1.72E-4 | 2.19E-9 |
| 0.9 | .696 | 6.95E-7 | 1.29E-4 | 4.98E-10 |
| high th. $\underline{\alpha} or \overline{\beta}$ | 1.8E-4 | 1.17E-5 | 6.09E-5 | 4.99E-6 |
| low th. $\underline{\alpha} or \overline{\beta}$ | 1.8E-4 | 1.8E-4 | 6.09E-5 | 6.09E-5 |

**Table 2: $\alpha$ and $\beta$ in NYT and PUBMED**

| Symbol | Description |
|---|---|
| $U, V$ | a collection of vectors, database |
| $J$ | join size |
| $n$ | database size, $|V| = n$ |
| $m$ | sample size |
| $u, v$ | vectors |
| $\tau$ | similarity threshold |
| $h$ | an LSH function, e.g. $h_1(u)$ |
| $k$ | # LSH functions for an LSH table |
| $\ell$ | # LSH tables in an LSH index |
| $g$ | $g = (h_1, \ldots, h_k)$, vector of LSH functions |
| $D_g, E_g$ | an LSH table using $g$ |
| $G$ | $G = \{g_1, \ldots, g_\ell\}$ |
| $I_G$ | an LSH index with using $G$ that consists of $D_{g_1}, \ldots, D_{g_\ell}$ |
| $B, C$ | a bucket in an LSH table, e.g. $B(u)$: the bucket of $u$ |
| $b_j, c_i$ | bucket counts, e.g. $b_j$ is the bucket count of bucket $B_j$ |
| T | **T**rue pairs, $(u, v)$ such that $sim(u, v) \geq \tau$ |
| F | **F**alse pairs, $(u, v)$ such that $sim(u, v) < \tau$ |
| H | **H**igh (expected) sim. pairs that are in the same bucket |
| L | **L**ow (expected) sim. pairs that are not in the same bucket |
| $P(T|H)$ | given $(u, v)$ s.t. $B(u) = B(v)$, prob. of $sim(u, v) \geq \tau$ |
| $P(H|T)$ | given $(u, v)$ s.t. $sim(u, v) \geq \tau$, prob. of $B(u) = B(v)$ |
| $P(T|L)$ | given $(u, v)$ s.t. $B(u) \neq B(v)$, prob. of $sim(u, v) \geq \tau$ |
| $P(L|T)$ | given $(u, v)$ s.t. $sim(u, v) \geq \tau$, prob. of $B(u) \neq B(v)$ |
| $S$ | stratum. e.g. $S_H$: set of pairs that are in the same bucket |
| $N$ | # pairs, e.g. $N_H$: # pairs in the same bucket |
| $M$ | # pairs in $V$, $M = \binom{|V|}{2}$ |

**Table 3: Summary of Notations**